\documentclass[a4paper,11pt]{article}
\usepackage{pos}

\title{The Belle II Upgrade Program}
\author*[a,b]{Peter Kri\v zan (for the Belle II Collaboration)}
\affiliation[a]{Faculty of Mathematics and Physics, University of Ljubljana,\\
  Jadranska 19, Ljubljana, Slovenia}

\affiliation[b]{Jo\v zef Stefan Institute,\\
  Jadranska 19, Ljubljana, Slovenia}

\emailAdd{peter.krizan@ijs.si}

\abstract{The Belle II detector at the SuperKEKB accelerator complex is covering a wide range of exciting physics topics. To achieve the project's research goals, a substantial increase of the data sample to 50~ab$^{-1}$ is needed, and for that, the luminosity has to reach the ambitious goal of $6 \times 10^{35}$ cm$^{-2}$ s$^{-1}$. The progress towards the design luminosity is accompanied by research and development of the accelerator, detector components, operation methods, as well as their upgrades. 
In the present contribution, we will discuss the status and plans of the project, timescales for upgrades, their motivations, and opportunities, an overview of upgrade options, and finish with an outlook and perspectives.}

\FullConference{%
  41st International Conference on High Energy Physics - ICHEP2022\\
  6-13 July, 2022\\
  Bologna, Italy
}

\begin{document}
\maketitle

\section{Introduction}

The Belle II detector~\cite{tdr,b2-nima} at the SuperKEKB accelerator complex~\cite{skb} is a super-B factory covering a wide range of exciting physics topics~\cite{b2pb}. To achieve the project's research goals, a substantial increase of the data sample to 50~ab$^{-1}$ is needed, and for that, the luminosity has to reach the ambitious goal of $6 \times 10^{35}$ cm$^{-2}$ s$^{-1}$. The progress towards the design luminosity is accompanied by research and development of the accelerator, detector components, operation methods, as well as their upgrades.  

In what follows, we will discuss the status and plans of the project, timescales for upgrades, their motivations, and opportunities, an overview of upgrade options, and finish with an outlook and perspectives.

\section{The SuperKEKB/Belle II program and plans}

The first phase of SuperKEKB operation in 2016, without the detector and collisions, was carried out to test the injection step and the two accelerator rings and for baking the 3~km of the accelerator vacuum chambers. This was followed by Phase 2 in 2018, with the first collisions with the complete accelerator but with an incomplete detector. Instead of the vertex detector, a dedicated background detector system known as Beast~2 was employed~\cite{Beast2}.

Physics runs started in Phase 3 (from 2019) luminosity runs with a complete detector, where the vertex detector was comprised of the pixel Detector (PXD) with a fully instrumented first layer and a partly instrumented layer 2, while the 4-layer micro-strip detector (SVD) was fully instrumented. The first physics paper appeared in January 2020~\cite{b2-first-physics-paper}.

The new accelerator presented several challenges, with an added operational complexity during the pandemic. It nevertheless reached a record peak luminosity of $4.7 \times 10^{34}$ cm$^{-2}$ s$^{-1}$. The path to reach the next milestone, a luminosity of $2 \times 10^{35}$ cm$^{-2}$ s$^{-1}$, has been identified. There are, however, still significant factors to achieve the target peak luminosity of $6 \times 10^{35}$ cm$^{-2}$ s$^{-1}$.

\begin{figure}[b]
\centering
\includegraphics[width=0.7\textwidth]{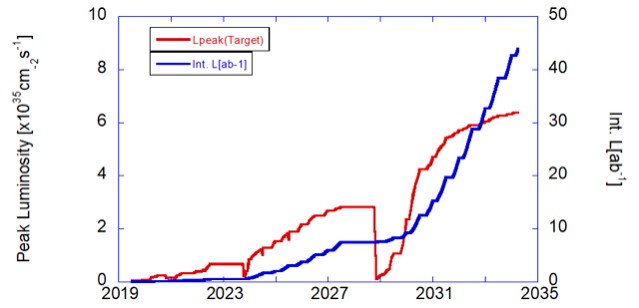}
\caption{Specific and integrated luminosity projection of SuperKEKB.
}
\label{fig:1}
\end{figure}

The path toward higher luminosity is a steep one. Challenges have been encountered in machine performance and stability coming from beam blow-up due to beam-beam effects, lower than expected beam lifetime, transverse mode coupling instabilities, low machine stability, injector capability, and aging infrastructure. 
Another set of challenges is background in the detector with single beam sources (beam-gas and Touschek scattering), luminosity-related (radiative Bhabha and two-photon processes), and injection background.

Mitigation measures include a consolidation of the accelerator complex and the detector. An 
international task force was established in 2020 to help with advice and ideas on how to consolidate the accelerator. Many countermeasures are under
development. To go beyond the luminosity of $\approx 2 \times 10^{35}$ cm$^{-2}$ s$^{-1}$, a major redesign of the Interaction
Region may be required. 

To consolidate the detector, the complete PXD detector has to be installed. Also, a part of the light sensors for one of the particle identification devices, the TOP (Time-of-Propagation) detector, have to be replaced by more robust devices. These two upgrades will be carried out in the Long Shutdown 1 (LS1) that started end of June 2022 and is expected to be finished by autumn 2023.  

A further upgrade of the detector is envisaged to make it more robust against backgrounds and improve its performance. This upgrade would be carried out in the  Long Shutdown 2 (LS2), expected to start by the end of 2026 or in 2027. 
LS2 is motivated by a (still to be defined) redesign of the interaction region (IR), with a replacement of the 	superconducting final quadrupoles. This is a window of opportunity for significant detector upgrades, including a possible replacement of the full vertex detector, the pixel (PXD) and silicon-strip (SVD) based parts. Although large uncertainties accompany the extent of accelerator upgrades in LS2, we must be ready and prepare technology choices for possible detector upgrades.  

On an even longer-term ($>2032$), options are discussed if a significant luminosity increase would become possible. A data sample of 
$\approx250$~ab$^{-1}$ would be interesting, although it is not clear at this time how to realize such a large increase in luminosity. A detailed study of the physics case is needed, and technology R\&D for an extreme-luminosity detector would have to start soon.

\section{Upgrades, main ideas, and time scale}

Crucial performance challenges that impact the physics reach are tracking at low momentum, vertex and interaction point resolution, calorimetry energy resolution and lepton identification, trigger efficiency, $K/\pi$ separation, and  $K_L$ detection.

\begin{figure}[b]
\centering
\includegraphics[width=0.7\textwidth]{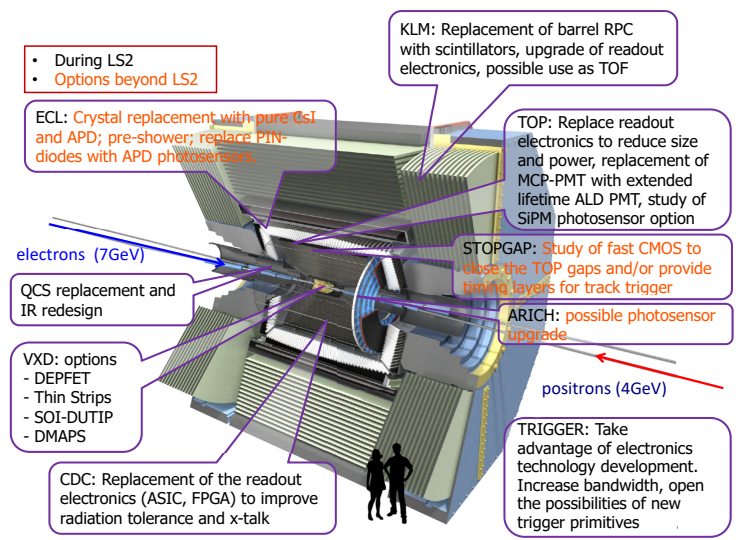}
\caption{Upgrades of the Belle II detector, as discussed for LS2 (black text) and LS3 (red text).
}
\label{fig:2}
\end{figure}

\subsection{Vertex detector (VXD) upgrade}

An upgrade of the vertex detector (VXD) is prepared to be aligned with a significant interaction region redesign, also allowing for large safety factors against background predictions. Possible performance improvements include an improvement of the impact parameter and vertexing resolution, an improved tracking performance for low transverse momentum ($p_T$) tracks, and a possible contribution to the L1 trigger.

Several proposals are under study and in the R\&D phase, taking advantage of recent technology developments. A thin and fine-pitch DSSD option assumes a 140~$\mu$m thin sensor with a z-pitch of $< 80$~$\mu$m, and a new ASIC for low noise operation. Another option assumes an upgraded DEPFET sensor with a higher radiation tolerance through higher gain and a faster read-out (few $\mu$s) with new ASICs, and a possible read-out re-orientation with respect to the present version of DEPFET.
In the option with silicon-on-insulator (SOI) pixels, the sensors are produced in the Lapis 200~nm process; the Dual Time pixel sensor (DuTiP) has a pitch of 45~$\mu$m, and a 2x60~ns integration time. Finally, a fully pixelated VXD concept is based on the CMOS-MAPS technology, produced in the Tower 180~nm process, an extension of the TJ-MONOPIX2 to the OBELIX sensor with a pitch of $< 40$~$\mu$m and 100~ns integration time; it assumes that the complete VXD detector is built of either all-Si modules or ALICE-ITS-like ladders.

\subsection{Central Drift Chamber (CDC)}
For the Central Drift Chamber (CDC), an upgrade of the read-out electronics is foreseen for LS2 with improved radiation tolerance and reduced cross-talk and power consumption. A new ASIC is under design, and a new FPGA and optical modules are being tested. 
Long-term studies are exploring options for sustaining higher rates and backgrounds expected at higher luminosities. For the present CDC, studies are underway with different gas mixtures. In the longer term, two further options are under study, one in which the inner part of the CDC is replaced by a silicon tracking device and one where a TPC tracker replaces the main part of the CDC with a Gridpix-like pixel read-out.

\subsection{Particle identification detectors}

{\bf TOP, Time-of-Propagation detector.} Here, the plan is to install life-extended Atomic Layer Deposition (ALD) micro-channel plate (MCP) PMTs to replace the standard MCP PMTs (in  LS1), and possibly to replace the standard ALD MCP PMTs (in LS2). A study of SiPMs as possible MCP PMT replacement is being pursued; it would require a dedicated cooling system and is thus possible only at a longer time scale (beyond LS2). A read-out electronics upgrade is under study where the present IRSX 8-channel 250~$\mu$m CMOS  ASIC  would be replaced by a TOPSoC 32-channel 130~$\mu$m CMOS, still under development; this new set-up would allow for a feature extraction inside ASIC, and reduced power consumption.

{\bf ARICH, aerogel RICH detector.} The detector is performing very well and has shown very little background susceptibility. Therefore no modifications are planned for the LS1 and LS2. Long-term studies include a photon detector upgrade with SiPMs or MCP-PMTs. For the SiPM option, irradiation tests are underway. Another promising candidate, a large area MCP-PMT (LAPPD, 20~cm$\times$20~cm active area), has been tested recently, and further tests are planned.  Two options are under study for the read-out electronics of the upgraded detector: a custom-developed version and the FASTiC chip (developed for the next upgrade of LHCb RICHes). The impact of a possible aerogel upgrade is under investigation.

{\bf STOPGAP detector.} This proposed device takes advantage of the development of fast CMOS sensors. Two ideas are discussed, one being to fill the gaps in the TOP detector with a $\approx 1$~cm$^2$ granularity to improve the kaon detection efficiency by about 10\% by covering the full solid angle. In the second proposal, two full timing layers are added at lower radii ($\approx 250$~mm, $\approx 450$~mm) to provide PID for low-momentum particles in the context of a larger VXD detector; such a detector could also provide trigger information. This interesting concept for longer-term upgrades requires more R\&D effort.

\subsection{Electromagnetic calorimetry}

The electromagnetic calorimeter performs very well despite of the larger background levels compared to the Belle experiment. Only possible long-term upgrades are therefore discussed. In one of the scenarios, the present scintillator crystals - CsI(Tl) - would be replaced by a considerably faster pure CsI scintillator, reducing the pile-up. A wavelength-shifting (WLS) window would  shift the wavelength of the  scintillation light to match the PiN diode sensitivity. In the second scenario, a pre-shower detector would be installed to reduce the background and pile-up and to determine photon direction and timing. The third option foresees a replacement of the present scintillation light sensors (PiN diodes) with APDs; this would improve the resolution and reduce the noise.
We note that all three options are complex and expensive and are thus only considered in long-term planning.

\subsection{Muon and $K_L$ detector: KLM}

In the muon detection system, a replacement of the RPC detectors with scintillator bars, WLS fibers, and  SiPMs as light sensors is discussed for the entire system in a similar way as already carried out for the endcaps and the first two layers in the barrel part of the spectrometer. By this, we would increase the rate capability of the system. This would be accompanied by a read-out electronics upgrade, with a more compact read-out and a possible data-push architecture. Possible use as a TOF detector is also discussed; a time resolution of around 100~ps would improve the $K_L$ identification. 
%Ongoing studies of scintillators and SiPM read-out arrangement for high time resolution

\subsection{Trigger}

The trigger system upgrade foresees installing a more powerful UT4 board for new CDC Front End electronics. Such an upgrade would allow for more bandwidth and the use of all CDC TDC and ADC information and would make many trigger improvements possible. Detailed technical documents on this upgrade are in preparation.

\subsection{Polarized electron beam}

An exciting possibility of further expanding the experiment's physics reach would be using a polarized beam~\cite{b2pb}. 
 The plan is to reach a 70\% polarization with an 80\% polarized source. New hardware for the polarization upgrade would include a low-emittance polarized source, spin rotators, and a Compton polarimeter to monitor the longitudinal polarization. Beam polarization options at SuperKEKB are under active study; there are some indications that such a modification may be possible on a more rapid timescale.

%\section{...}

\section{Summary}

Belle II and SuperKEKB have started a successful physics run. 
Machine improvements are being studied and implemented to reach the target luminosity. Detector upgrade ideas are being explored, and R\&D is in progress to achieve better robustness against background and radiation damage, improve the physics performance, and be ready for a possible interaction region redesign. 

The Belle II upgrade organization is in place. The Upgrade Working Group and Upgrade Advisory Committee have been set up to 	help establish priorities and direct the effort. A Belle II Upgrades White paper has been submitted to the US Snowmass process.\cite{b2-upgrade-snowmass} The transition to a construction project is needed soon. While the SuperKEKB International Task Force is finalizing its recommendations, the preparation of an Upgrade Conceptual Design Report has started with the aim to be ready end of 2023. 

For a longer-term perspective of the Belle II experiment, it is important to start exploring the options. It is worth noting that there is  considerable physics potential at even higher luminosities and with an even better detector.

\end{document}